# Properties of Assembly of Superparamagnetic Nanoparticles in Viscous Liquid

N. A. Usov[1,2*], R. A. Rytov[1] and V.A. Bautin[1]

[1]*National University of Science and Technology «MISiS», 119049, Moscow, Russia*

[2]*Pushkov Institute of Terrestrial Magnetism, Ionosphere and Radio Wave Propagation, Russian Academy of Sciences, IZMIRAN, 108480, Troitsk, Moscow, Russia*

**Abstract.** Detailed calculations of the specific absorption rate (SAR) of a dilute assembly of iron oxide nanoparticles with effective uniaxial anisotropy dispersed in a liquid are performed depending on the particle diameters, the alternating (ac) magnetic field amplitude and the liquid viscosity. For small and moderate ac magnetic field amplitudes $H_0$ with respect to particle anisotropy field $H_k$ the SAR of the assembly as a function of the particle diameter passes through a characteristic maximum and then reaches a plateau, whereas for sufficiently large amplitudes, $H_0 \sim H_k$, the SAR increases monotonically as a function of particle diameter. This difference is a consequence of realization of viscous and magnetic oscillation modes for particle unit magnetization vector and director for moderate and sufficiently large $H_0$ values, respectively. It is found that the SAR of the assembly changes inversely with the viscosity only in a viscous mode, for nanoparticles of sufficiently large diameters. In the developed magnetic mode the SAR of the assembly is practically independent of the viscosity, since in this case the nanoparticle director only weakly oscillates around the ac magnetic field direction. At moderate amplitudes of the ac magnetic field the SAR values of the assembly in the liquid and in the solid matrix are found to be close, except of the range of large particle diameters and sufficiently low viscosity. However, at large field amplitudes the SAR of randomly oriented assembly of nanoparticles in a solid matrix is approximately two times less than that in a liquid, because a significant fraction of nanoparticles of the assembly in the solid matrix is not optimally oriented with respect to the ac magnetic field direction. The conditions for the validity of the linear response theory have been clarified by comparison with the numerical simulation data.

*Corresponding author           N.A. Usov (usov@obninsk.ru)

## 1. Introduction

The behavior of an assembly of magnetic nanoparticles in a static or alternating (ac) magnetic field is of great interest due to the important applications of these assemblies in biomedicine [1]. In particular, the ability of magnetic nanoparticles to effectively absorb the energy of ac magnetic field and dissipate it in the environment can be used in magnetic hyperthermia, a new promising method for treating cancer [2-5]. Generally speaking, for applications in magnetic hyperthermia it is necessary to study the specific absorption rate (SAR) of an assembly in a biological environment. However, in a biological medium there is an intense interaction of nanoparticles with intracellular structures [6,7] and the formation of clusters of particles [8,9], which leads to a significant increase in the intensity of magnetic dipole interaction in the assembly [10,11]. In addition, the rheological properties of biological media have not yet been fully studied [12]. All these factors significantly complicate the analysis of experimental results obtained for nanoparticles in a biological medium. Therefore, it is a common practice to carry out SAR measurements of prepared assemblies in a liquid since it helps to quickly assess the prospects of a particular assembly for use in magnetic hyperthermia.

To date, a large number of SAR measurements have been carried out in liquids of various viscosities [6,7,13-21]. In a number of cases [13-15], very high SAR values, more than 1 kW/g, were obtained at high frequencies, $f = 300 - 700$ kHz, and large amplitudes of the ac magnetic field, $H_0 = 300 - 600$ Oe. However, for medical reasons one should strive to obtain a sufficient thermal effect at moderate values of the frequency and magnetic field amplitude. In any case, the product $f \cdot H_0$ must satisfy the Brezovich [22], or at least Hergt et al. [23] criterion, $f H_0 \leq 6.28 \times 10^4$ kHz*Oe. Besides, the question about the influence of the liquid viscosity, magnetic parameters and characteristic sizes of nanoparticles [13-21] on the SAR of the assembly requires a more detailed study.

An early paper [13] stated that the linear response theory (LRT) developed by Rosensweig [24] describes the experimental SAR measurements in a viscous liquid very satisfactorily. Later, however, it was experimentally discovered [15, 16] that significant deviations from the LRT predictions arise with an increase in the ac magnetic field amplitude. It should be noted that theoretical calculations of SAR of dilute assemblies of magnetic nanoparticles in a viscous liquid are complicated by the fact that a magnetic nanoparticle in a liquid can rotate as a whole under the action of fluctuating thermal torques and external ac magnetic field. The spatial orientation of a single-domain nanoparticle with uniaxial anisotropy can be specified by a unit vector, the so-called director $n$, which indicates the easy axis direction of the particle's magnetic anisotropy. Furthermore, the directions of the particle unit magnetization vector $\alpha$ and the director $n$ are not independent, since the magnetic anisotropy energy of the particle depends on the angle between these vectors. In the LRT [24] a number of simplifying assumptions were introduced that allow one to gracefully bypass a detailed theoretical analysis of physical processes leading to the absorption of the energy of ac magnetic field by an assembly of nanoparticles in a liquid. However, the area of applicability of the LRT is still poorly studied.

In subsequent works [25-30], various generalizations of LRT were developed, based on the introduction of the expressions for the dynamic magnetic susceptibility or for the relaxation times of superparamagnetic nanoparticles in a viscous liquid. An alternative approach was also developed [31-36] based on the use of a system of stochastic equations to describe the dynamics of the unit magnetization vector and the particle director at a finite temperature. A useful overview of the current state of the problem can be found in Ref. 37.

To study the behavior of a dilute assembly of magnetic nanoparticles in a viscous liquid, a system of stochastic equations that describes the coupled dynamics of vectors $\alpha$ and $n$ under the action of an external magnetic field and thermal fluctuations in the medium was proposed [31]. Two vibration modes (the so called viscous and magnetic) were shown to be realized for vectors $\alpha$ and $n$ depending on the amplitude $H_0$ of ac magnetic field compared to the particle anisotropy field $H_k$, In the viscous mode, which is realized at low and moderate amplitudes, $H_0 < H_k/2$, the vectors $\alpha$ and $n$ move approximately in unison. A magnetic mode arises with an increase in the field amplitude, $H_0 \sim H_k$. In magnetic mode the particle director experiences only small oscillations about the magnetic field direction, whereas the magnetic vector periodically jumps between equivalent potential wells with the frequency of the external alternating field.

The orientation of nanoparticle directors in a sufficiently strong ac magnetic field was already observed in the early experiment [14]. The existence of the magnetic oscillation mode has also been confirmed in recent experiments [38,39]. Meanwhile, the results of SAR measurements of nanoparticle assemblies distributed in a liquid are usually compared [13,15,16,28-30] with the LRT predictions, although formally this theory can be valid only in a relatively weak magnetic field.

In this work, using the previously developed approach [31], detailed calculations of the SAR of a dilute assembly of iron oxide nanoparticles are performed depending on the particle diameters, the alternating magnetic field amplitude and the liquid viscosity. The conditions for the applicability of the LRT have been clarified. The SAR of dilute assemblies distributed in a solid matrix and in a viscous liquid has been compared.

## 2. Results and Discussion

### a. Dilute nanoparticle assembly in viscous liquid

A behavior of a dilute assembly of magnetic nanoparticles in a viscous liquid can be studied [31] by solving a set of stochastic equations that describe the coupled dynamics of the unit magnetization vector $\alpha$ and the director $n$ of a particle in ac magnetic field. A brief description of this approach is given in the methods section. Using this technique in this work we investigate in detail the dependence of the SAR of a dilute assembly of quasi-spherical single-domain iron oxide nanoparticles with a saturation magnetization $M_s = 350$ emu/cm$^3$ and an effective uniaxial anisotropy constant $K_1 = 10^5$ erg/cm$^3$ on the particle size in the range of diameters $D = 15 - 55$ nm, on the amplitude of the ac magnetic field, $H_0 = 50 - 500$ Oe, and the kinematic viscosity of the liquid, $\eta = 0.01 - 0.2$ g/(cm·s). The density of magnetic nanoparticles is taken to be $\rho = 5.0$ g/cm$^3$, the temperature of the system $T = 300$ K. The parameters mentioned are of greatest interest for the application of magnetic nanoparticles in magnetic hyperthermia [2-5]. The calculations were carried out at magnetic field frequency $f = 300$ kHz, that is a typical value for SAR measurements in a viscous liquid [13-19]. The results of the calculations performed are shown in Fig. 1.

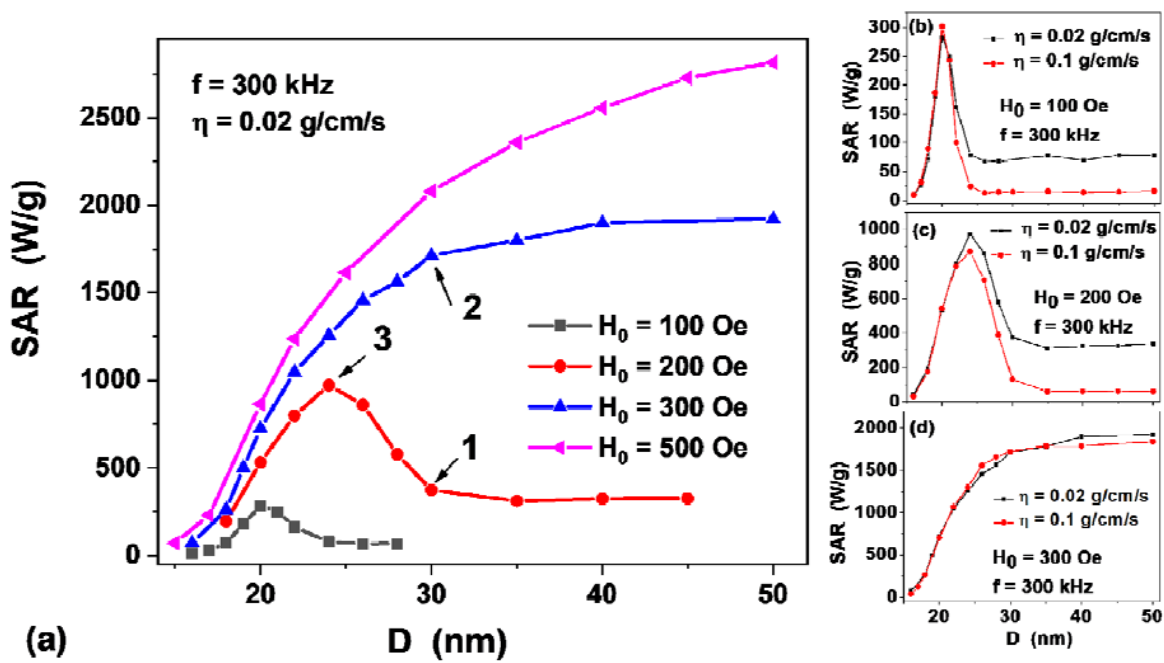

Fig. 1. a) SAR of a dilute assembly of magnetic nanoparticles depending on the particle diameter for different ac magnetic field amplitudes $H_0$ at the liquid viscosity $\eta = 0.02$ g/(cm·s). b)-d) dependence of SAR on the liquid viscosity for various ac magnetic field amplitudes.

In Fig. 1a one notes a significant difference in the dependence of SAR on the particle diameter for the cases of moderate, $H_0 = 100, 200$ Oe, and sufficiently large amplitudes, $H_0 = 300, 500$ Oe, of the ac magnetic field. In the first case the SAR as a function of the particle diameter passes through a characteristic maximum and then reaches a plateau, while in the second case it increases monotonically with an increase in the particle diameter. As noted earlier [31], this difference is due to the fact that in the range of amplitudes $H_0 < H_k/2$, where $H_k = 2K_1/M_s = 571$ Oe is the particle anisotropy field, the dynamics of the unit magnetization vector $\boldsymbol{\alpha}$ and director $\boldsymbol{n}$ occurs mainly in the so called viscous oscillation mode, where the movement of both vectors is highly correlated. At the same time, for sufficiently large amplitudes, $H_0 \sim H_k$, a magnetic oscillation mode is realized. In the magnetic mode, in the course of time the particle directors are oriented along the line of action of the ac magnetic field, while the magnetic vector periodically jumps between two equivalent magnetic potential wells. In some cases, there is also an intermediate regime in which the viscous and magnetic motion of the vectors alternate.

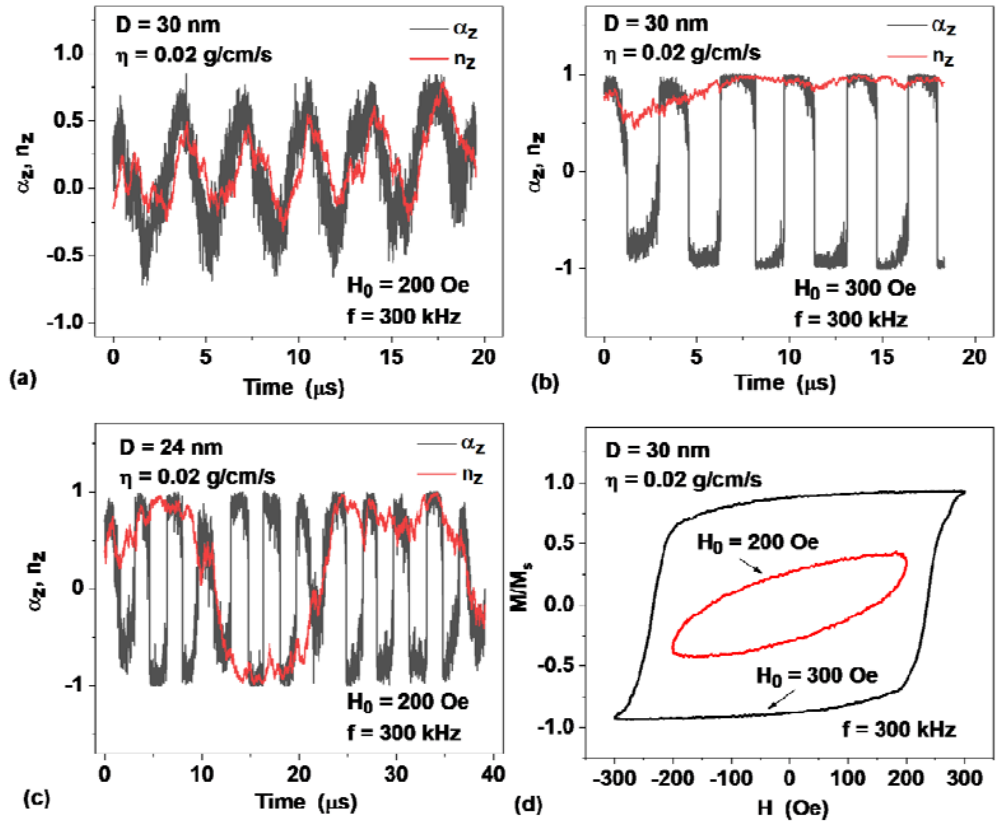

Fig. 2. The nature of the motion of the unit magnetization vector and director of an individual nanoparticle of the assembly in a viscous liquid for different cases: a) developed viscous mode, point 1 in Fig. 1a, b) developed magnetic mode, point 2 in Fig. 1a, c) intermediate regime, point 3 in Fig. 1a. d) low frequency hysteresis loops for viscous and magnetic modes at points 1 and 2 in Fig. 1a, respectively.

Fig. 2a shows the nature of the motion of vectors $\boldsymbol{\alpha}$ and $\boldsymbol{n}$ in the viscous mode. Here, the black and red curves show the projections of the unit magnetization vector and the director of an individual, randomly selected nanoparticle of the assembly onto the applied magnetic field direction for several periods of field variation. This type of motion of vectors $\boldsymbol{\alpha}$ and $\boldsymbol{n}$ is characteristic when the SAR of the assembly reaches a plateau. In particular, Fig. 2a shows the oscillations of the z-components of these vectors at $H_0 = 200$ Oe, $f = 300$ kHz, $D = 30$ nm, which corresponds to point 1 in Fig. 1a. However, with an increase in the ac field amplitude, $H_0 = 300$ Oe (point 2 in Fig. 1a), the magnetic mode occurs in the same assembly. As Fig. 2b shows, in the magnetic mode the longitudinal projection of the director is close to $n_z = +1$ (the value $n_z = -1$ is equally possible), whereas projection of the unit magnetization vector oscillates between the values $\alpha_z = \pm 1$ with the frequency of the ac magnetic field. It should be noted that for particles of large diameters, and in liquid with appreciable viscosity, the establishment of stationary motion in the magnetic mode can occur rather slowly, over several tens of periods of the ac magnetic field.

Finally, Fig. 2c shows the intermediate mode of motion of vectors $\boldsymbol{\alpha}$ and $\boldsymbol{n}$ at point 3 in Fig. 1a. In this regime, the longitudinal projection of the director also periodically jumps between the values $n_z = \pm 1$, but the frequency of this jump is much lower than the frequency $f$ of the applied magnetic field. Fig. 2d shows the low frequency hysteresis loops of a dilute assembly consisting of $N_p = 400$ nanoparticles with a diameter $D = 30$ nm. These loops correspond to the viscous and magnetic modes of the assembly at points 1 and 2 in Fig. 1a, respectively. It is easy to see that the shape of the hysteresis loop for the magnetic mode is close to that of an ideal rectangular hysteresis loop. Note, that for rectangular hysteresis loop the SAR of the assembly has the maximum possible value, SAR = $4M_sH_0f/\rho$, [40].

It is interesting to note that at point 3 in Fig. 1a the SAR of the assembly peaks as a function of the particle diameter. For the case of $H_0 = 200$ Oe the SAR at the maximum, at $D = 24$ nm, is 972.5 W/g, while the characteristic SAR values at the plateau, at $D > 30$ nm, are only 320 - 350 W/g. Similarly, for the case of $H_0 = 100$ Oe, the maximum SAR at $D = 20$ nm is 283 W/g, while the SAR at the plateau, $D > 24$ nm is approximately 70 W/g. Consequently, to generate heat in a viscous liquid at moderate amplitudes of ac magnetic field it is preferable to use assemblies of iron oxide nanoparticles with characteristic diameters $D = 20 - 24$ nm. Of course, with an increase in the field amplitude up to $H_0 = 300$ Oe much higher SAR values can be obtained at the same frequency $f = 300$ kHz, on the order of 1600 - 1700 W/g, for nanoparticle diameters $D > 30$ nm. But the use of variable magnetic fields of large amplitude is energetically expensive and unsafe in a medical clinic. Besides, according to the Hergt et al. criterion [23], the product $f \cdot H_0$ should be limited for medical reasons.

Let us now turn to the question of the influence of liquid viscosity on the SAR values. A common place is the statement [6,13,16,18] that the SAR of an assembly in a liquid decreases with an increase in the liquid viscosity. However, as Figs. 1b-1d, show, at least for a dilute assembly, neglecting the clustering of the particles, this statement requires a significant correction. The SAR of the assembly does indeed changes inversely with the viscosity, but only in the developed viscous mode, when it reaches a plateau with an increase in the nanoparticle diameter. On the other hand, as Figs. 1b-1c show, an increase in the viscosity has only a weak effect on the SAR value at the maximum, in the viscous and intermediate mode, at moderate field amplitudes, $H_0 < H_k/2$. In addition, according to Fig. 1d, in the developed magnetic mode the SAR is practically independent of the viscosity, since in this case the nanoparticle director only weakly oscillates around the ac magnetic field direction.

**b. Effective relaxation time of nanoparticle in liquid**

It has recently become popular to apply LRT [24] to interpret the data of the SAR measurements for assemblies distributed in a viscous fluid. In [24] the analytical expression for the SAR of an assembly was formally obtained in the limit of a weak ac magnetic field

$$SAR = \chi_0 \frac{\omega \tau_{ef}}{1 + (\omega \tau_{ef})^2} \frac{f}{\rho} H_0^2 . \qquad (1)$$

Here $\chi_0 = \pi M_s^2 V/3k_B T$ is the initial magnetic susceptibility of an assembly of superparamagnetic nanoparticles [41], $V = \pi D^3/6$ is the nanoparticle volume, $k_B$ is the Boltzmann constant, and $\omega = 2\pi f$ is the circular frequency. Eq. (1) contains also the characteristic time of magnetic relaxation of the assembly $\tau_{ef}$. This quantity was taken [24] to be equal to the Shliomis relaxation time [42]

$$\tau_{ef} = \frac{\tau_B \tau_N}{\tau_B + \tau_N}, \qquad (2)$$

where $\tau_B = 3\eta V/k_B T$ is the time of the Brownian orientational relaxation of the particle directors under the action of thermal fluctuations in the liquid, obtained for polar liquids by Debye [43]. Note, that in the presence of a nonmagnetic layer on the particle surface the particle volume $V$ in this formula should be replaced by the effective volume $V_\eta$. However, such a replacement does not lead to qualitatively new results; therefore, in this work we assume for simplicity $V_\eta = V$. Further, $\tau_N$ is the Néel relaxation time of the average magnetic moment of an assembly of immobile superparamagnetic particles. For this quantity Brown [44] obtained an analytical approximation in the limit of a sufficiently large reduced energy barrier, $R_b = K_1 V/k_B T \gg 1$

$$\tau_N = \frac{1}{f_0} \exp\left(\frac{K_1 V}{k_B T}\right); \qquad f_0 = \frac{4\kappa \gamma_1 K_1}{M_s} \sqrt{\frac{K_1 V}{\pi k_B T}}. \qquad (3)$$

Here $\kappa$ is the magnetic damping constant, $\gamma_1 = \gamma/(1+\kappa^2)$, $\gamma$ is the gyromagnetic ratio.

Obviously, the relaxation of the average magnetic moment of an assembly in a liquid occurs due to the simultaneous action of two processes, namely, due to the motion of the unit magnetization vectors of the particles with respect to the directions of the easy anisotropy axes, and due to the rotation of particles in the liquid as a whole. According to the Shliomis hypothesis [42], both of these processes occur independently of each other.

Although the assumption of the independence of relaxation processes is not obvious, the Shliomis hypothesis is widely used in the analysis of experimental results [13,16,27-30]. In [26] the Shliomis relation was derived by an approximate solution of the Fokker – Planck equation in the limit of a weak external magnetic field. Due to the importance of this hypothesis, in this work the process of relaxation of a dilute assembly of nanoparticles in a liquid is also studied using numerical simulation. For this, let us prepare the assembly at time $t = 0$ in the initial ordered state, when the directors and unit magnetization vectors of all particles are parallel to the z-axis of the Cartesian coordinates, $\mathbf{n}_i = \mathbf{\alpha}_i = (0,0,1)$, $i = 1, 2, ..$ , $N_p$. The time evolution of the assembly at $t > 0$ can be studied by solving the system of stochastic equations for the motion of vectors $\mathbf{\alpha}$ and $\mathbf{n}$ given in the methods section. The average reduced magnetic moment of the assembly and the average moment of the particle directors during the evolution of the assembly are calculated as follows

$$\langle \vec{m}(t) \rangle = \frac{1}{N_p} \sum_{i=1}^{N} \vec{\alpha}_i; \qquad \langle \vec{n}(t) \rangle = \frac{1}{N_p} \sum_{i=1}^{N} \vec{n}_i. \qquad (4)$$

The last vector characterizes the change in the total spatial orientation of the particles over time. Obviously, as a result of the relaxation of a dilute assembly of superparamagnetic nanoparticles to the equilibrium state, the average values of both vectors should be equal to zero, <$\mathbf{m}$> = <$\mathbf{n}$> = 0.

The calculation of relaxation processes in a liquid was performed for a dilute assembly with the same magnetic parameters, $M_s$ = 350 emu/cm$^3$, $K_1$ = 10$^5$ erg/cm$^3$, for particles of various diameters. To obtain statistically reliable results the calculations were carried out for a sufficiently large assemblies of particles, $N_p$ = 800 - 1000. The numerical time step in these calculations was chosen to be 1/30 of the characteristic precession time of unit magnetization vectors $T_p$, since it was shown earlier [45] that such a choice guarantees sufficient accuracy of the numerical calculations.

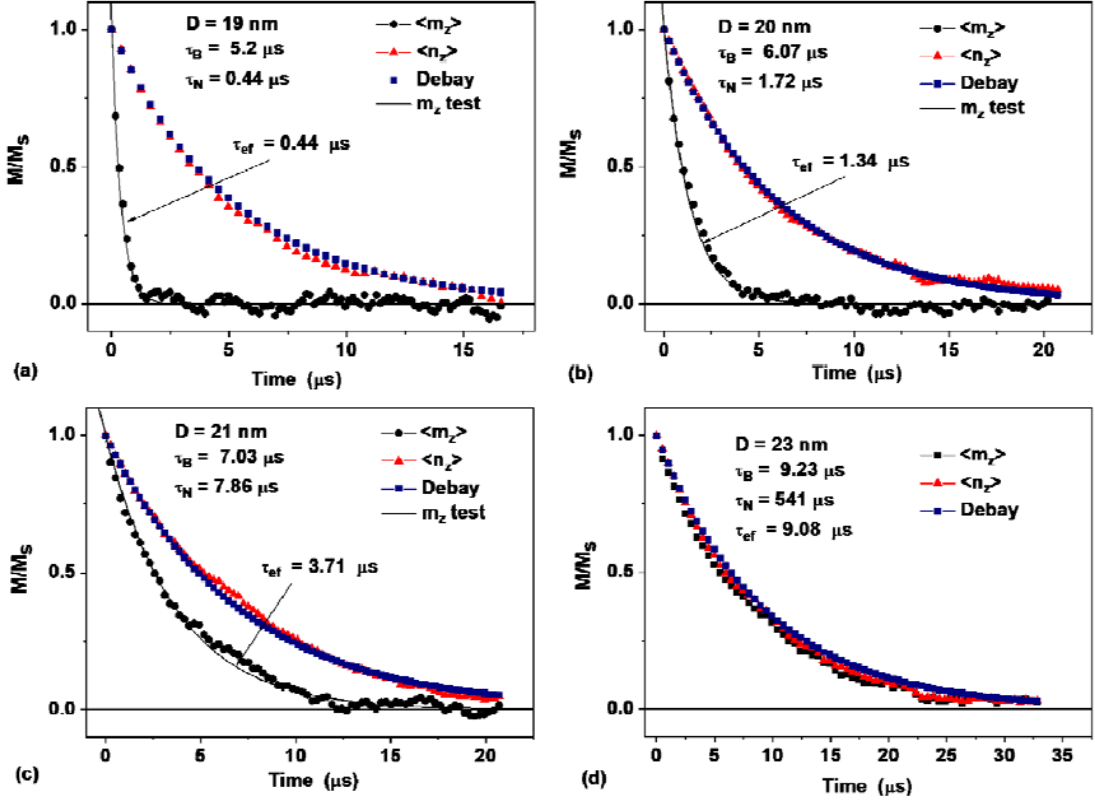

Fig. 3. Relaxation curves of dilute assemblies of magnetic nanoparticles in a liquid with viscosity $\eta = 0.02$ g/(cm·s) at a temperature $T = 300$ K for particles of different diameters: a) $D = 19$ nm, b) $D = 20$ nm, c) $D = 21$ nm and d) $D = 23$ nm. Circular and triangular dots show the time dependence of the projections $<m_z(t)>$ and $<n_z(t)>$, respectively. The square dots show the Debye orientation relaxation process with the corresponding relaxation time $\tau_B$. The solid curves show the exponential approximation of the $<m_z(t)>$ dependence with the corresponding effective relaxation time $\tau_{ef}$, Eq. 2.

In Figs. 3a-3d black circular dots show the time dependence of the z-component of the average reduced magnetic moment of the assembly, $<m_z(t)>$, for nanoparticles of various diameters, $D = 19, 20, 21$ and 23 nm, respectively. The red triangular dots show the time evolution of the component $<n_z(t)>$ that characterizes the average spatial orientation of nanoparticles of the assembly. For comparison, the blue square dots show the rate of orientational relaxation in a liquid of assembly of nonmagnetic nanoparticles of the corresponding diameter with the Debye relaxation time $\tau_B$. As seen from Fig. 3, the curves of orientational relaxation of the assembly in all cases very closely follow the corresponding Debye relaxation curves. Thus, the presence of magnetic degrees of freedom of nanoparticles has practically no effect on the process of orientational relaxation of the assembly in a viscous liquid.

On the other hand, the presence of orientational degrees of freedom of magnetic nanoparticles significantly affects the relaxation process of the average magnetic moment of the assembly. As Figs. 3a-3c show, the magnetic relaxation curves are very closely approximated by exponents with the Shliomis relaxation time, Eq. (2). These approximations are shown in Fig. 3a-3c with solid lines. Similar results were obtained for dilute assemblies of iron oxide nanoparticles in liquids of different viscosities, $\eta = 0.01 – 0.2$ g/cm/s. Thus, the Shliomis hypothesis is fully justified in the absence of applied magnetic field. It is reasonable to assume that it remains valid in a sufficiently weak static or ac magnetic field.

As Fig. 3 shows, for given magnetic parameters and liquid viscosity the relaxation times $\tau_N$ and $\tau_B$ turn out to be nearly equal for particles with diameter $D \approx 21$ nm. Since the relaxation time $\tau_N$ grows exponentially as a function of particle size, the ratio $\tau_N/\tau_B$ becomes very large already at $D = 23$ nm. Then it follows from Eq (2) that the effective relaxation time $\tau_{ef} = \tau_B$. Thus, for particles with diameters $D \geq 23$ nm, the magnetic relaxation occurs only due to the rotation of particles in the liquid as a whole. Indeed, as Fig. 3d shows, for the case $D = 23$ nm the relaxation curves $<m_z(t)>$ and $<n_z(t)>$ practically coincide with each other and with the Debye relaxation curve.

On the other hand, for particles with diameters $D \leq 19$ nm, the Néel – Brown relaxation time (3) is much shorter than the Debye relaxation time, therefore $\tau_{ef} = \tau_N$. Accordingly, in this range of diameters the magnetic relaxation of the assembly occurs due to the Néel - Brown mechanism. The slow orientational relaxation for nanoparticles of such diameters is insignificant. As Fig. 3 shows, the phenomenological Eq. (2) also describes well the time dependence of magnetic relaxation in a narrow range of nanoparticle diameters, 19 nm $\leq D \leq$ 23 nm, in which the relaxation regime changes.

Note that under the assumption that the relaxation processes under consideration are independent, a qualitative explanation of the validity of Eq. (2) can be obtained as follows. Let us assume that without taking into account orientational relaxation, the process of magnetic relaxation proceeds according to the law $m_z(t) = m_z(0)\exp(-t/\tau_N)$. But due to orientational relaxation, the projection of the average magnetic moment of the assembly on the selected axis decreases according to the equation $m_z \sim \exp(-t/\tau_B)$. As a result of the simultaneous and independent action of both processes, one obtains the time dependence $m_z(t) \sim \exp(-t/\tau_B)\exp(-t/\tau_N) \sim \exp(-t/\tau_{ef})$.

**c. Applicability of linear response theory**

As noted in the introduction, at the moment it is not clear what it the range of validity of the Eqs. (1), (2). In this section the dependence of the SAR of the assembly on the amplitude of the ac magnetic field is investigated for magnetic nanoparticles of various diameters. It was shown in the previous section that for given particle magnetic parameters and liquid viscosity, the effective relaxation time of the assembly is determined by the Néel mechanism for particles with diameter $D \leq 19$ nm, and by the Debye mechanism for $D > 22$ nm. One may expect therefore a different SAR behavior in these intervals of particle diameters.

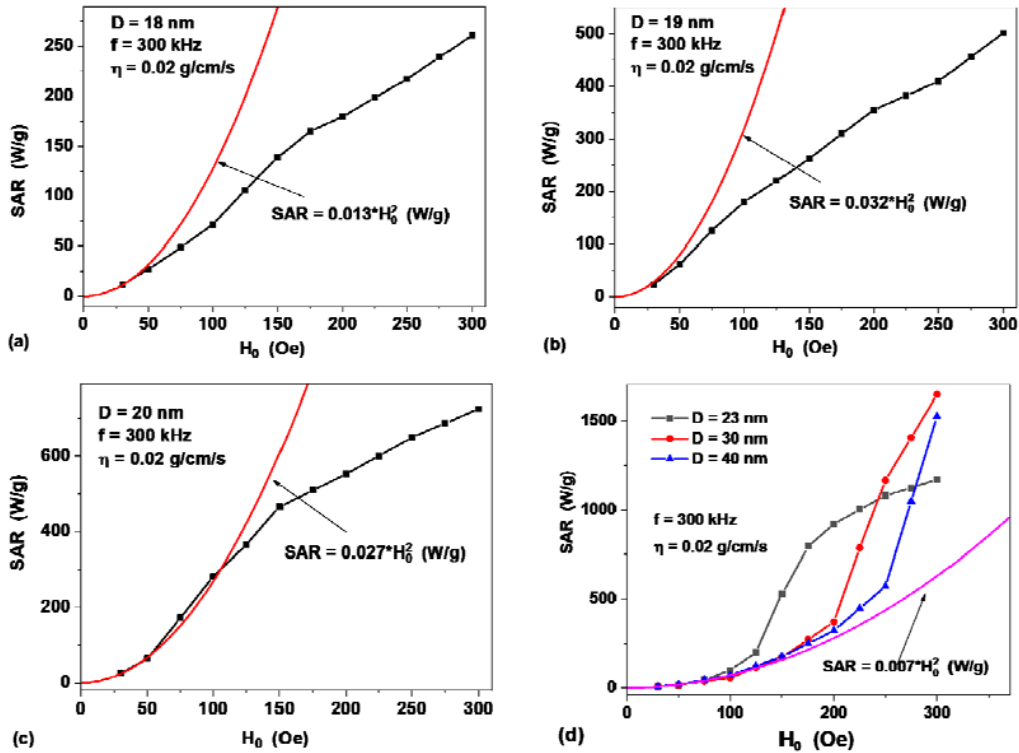

Fig. 4. Dependence of the SAR of a dilute assembly in a liquid (dots) on the magnetic field amplitude for particles of different diameters: a) $D = 18$ nm, b) $D = 19$ nm, c) $D = 20$ nm and d) $D = 23$-$40$ nm. The solid curves are drawn in accordance with Eqs. (1) - (3).

As Figs. 4a, 4b show, for nanoparticles with diameters $D \leq 19$ nm, the linear regime is limited by the value $H_0 = 50$ Oe. Note that for small particle diameters, $\omega \tau_{ef} \ll 1$, when SAR is determined by the Néel mechanism, one obtains from Eq. (1)

$$SAR = \frac{2\pi^2}{3} \frac{M_s^2}{k_B T \rho} V \tau_N f^2 H_0^2. \qquad (5)$$

Thus, in this limit the SAR of the assembly does not depend on the liquid viscosity. However, there is an exponential dependence on the reduced energy barrier and a quadratic dependence on the ac field frequency.

According to Figs. 4c, 4d, with an increase in the particle diameter the linear regime expands up to $H_0 = 200$ Oe for particles with $D = 40$ nm. Note that for particles with large diameters, when the condition $\omega \tau_{ef} \gg 1$ is satisfied, Eq. (1) takes the form

$$SAR = \frac{M_s^2}{18 \eta \rho} H_0^2. \qquad (6)$$

Therefore, in this area the SAR of the assembly is inversely proportional to the liquid viscosity. In addition, it does not depend on either the particle diameter or the ac field frequency. Numerical data in Fig. 4d confirm the weak dependence of SAR on particle diameters in this limit.

It is interesting to note that according to Figs. 4a-4c, with the development of the nonlinear regime the SAR values obtained by numerical simulation for particles of small diameters, $D \leq 20$ nm, decrease in comparison with the LRT predictions. The opposite situation is observed in Fig. 4d for large particle diameters, $D \geq 23$ nm.

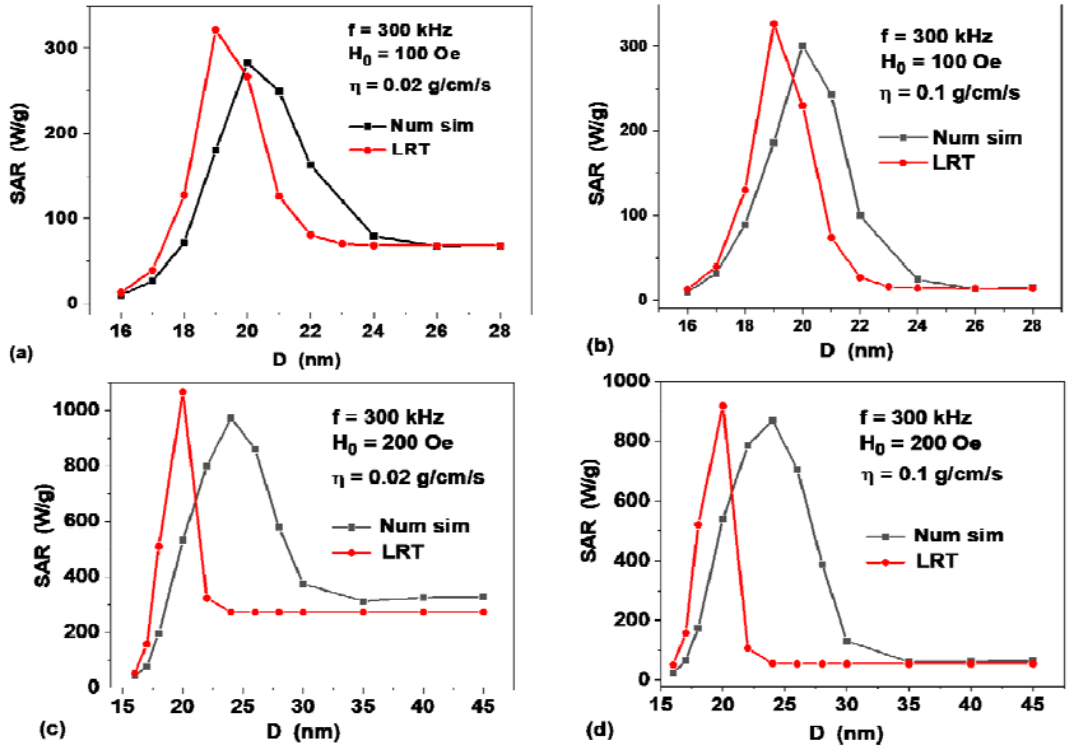

Fig. 5. Comparison of the numerical simulation data and the LRT approximation for different values of the liquid viscosity and the moderate values of ac magnetic field amplitude: a)-b) $H_0 = 100$ Oe, c)-d) $H_0 = 200$ Oe.

Fig. 5 shows a comparison of the numerical simulation data with the LRT results depending on the particle diameters at different values of $H_0$ and $\eta$. One can see that numerical data are sufficiently close to LRT only for small particle diameters, $D \leq 16$ nm, when the assembly SAR is determined by the Néel mechanism, and in the range of large diameters, $D \geq 24 - 30$ nm, where the viscous friction in a liquid dominates. In the intermediate range of diameters the absorption peak in the LRT is significantly shifted to small particle diameters, so that the difference in SAR values obtained by the two methods is very large, especially for $H_0 = 200$ Oe. At even higher amplitudes of ac magnetic field the LRT is not applicable.

The difference between the numerical results and LRT shown in Fig. 5 for the SAR of a dilute assembly in a viscous fluid is expected, since LRT is applicable, strictly speaking, under condition $H \ll H_k$ only. Indeed, Eq. (1) uses an expression for the initial magnetic susceptibility of the assembly, which is valid only in a weak magnetic field [41]. In addition, Eq. (2), which is essentially used in Eq. (1), characterizes the relaxation processes only in a weak magnetic field. It does not take into account the presence of magnetic mode which develops at sufficiently large amplitudes of the ac magnetic field.

**d. SAR of nanoparticles distributed in a liquid and in a solid matrix**

It is instructive to compare the SAR for dilute nanoparticle assemblies dispersed in different media, namely, a solid matrix and a viscous liquid. The magnetic parameters of the particles and the ac field frequency remain the same, the liquid viscosity varies within the range $\eta = 0.01 – 0.1$ g/(cm·s).

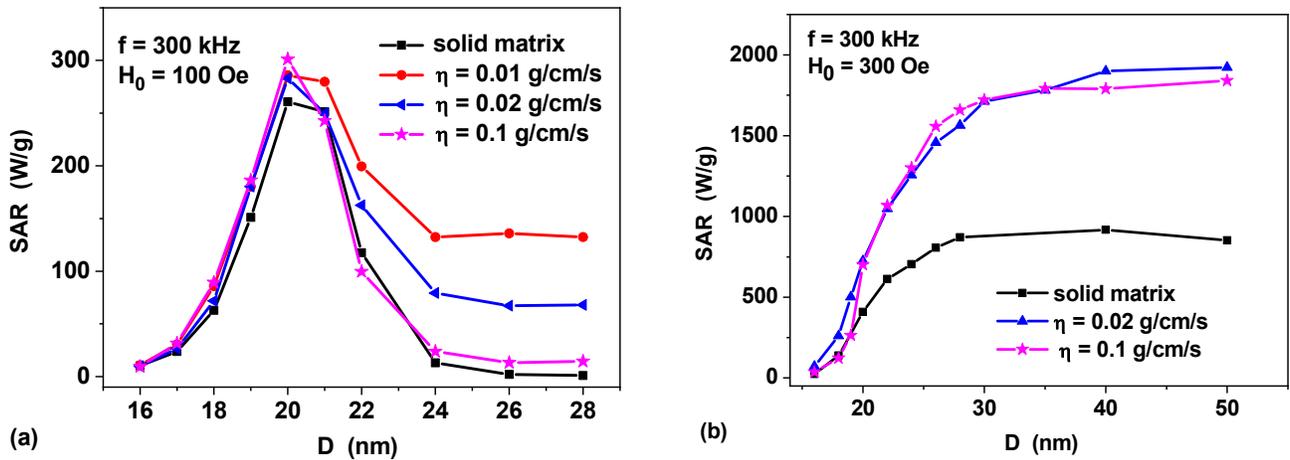

Fig. 6. Comparison of SAR of dilute assemblies of nanoparticles distributed in a solid matrix and in liquids with various viscosity, $\eta = 0.01 – 0.1$ g/(cm·s) for different amplitudes of ac magnetic field: a) $H_0 = 100$ Oe, b) $H_0 = 300$ Oe.

In Figs. 6a, 6b, black square dots show the dependence of SAR on particle diameter for randomly oriented assembly of nanoparticles distributed in a solid matrix. Other curves show a similar SAR dependence for assemblies of nanoparticles distributed in liquids of various viscosities. For a dilute randomly oriented assembly of uniaxial nanoparticles in solid matrix it was theoretically shown [40] that there is an optimal range of particle diameters where SAR reaches maximum at moderate amplitudes of ac magnetic field. Indeed, for particles of small diameters, the effect of thermal fluctuations of magnetic moments is predominant, so that the low-frequency hysteresis loops of such nanoparticles practically do not open. With an increase in the particle diameter the area of the frequency hysteresis loop increases, but for nanoparticles of sufficiently large diameters, having increased coercive force, it again decreases, since the conditions for the magnetization reversal of such particles at $H_0 \ll H_k$ are not satisfied.

It is interesting to note that as Fig. 6a shows, at moderate amplitudes of the ac magnetic field, $H_0 \ll H_k$, the SAR values of the assembly in the liquid and in the solid matrix are close. For small particle diameters in both cases the Néel relaxation mechanism dominates. Moreover, the maximum SAR values in both cases correspond to a narrow range of particle diameters, $D$ = 19 - 22 nm. The different SAR behavior arises only for $D$ > 22 nm, where the SAR of the assembly in the solid matrix rapidly tends to zero, since in this area the particle effective energy barriers are high and the probability of the magnetization reversal is small. At the same time, as Fig. 6a shows, for assembly of nanoparticles in a liquid with a sufficiently low viscosity, $\eta$ = 0.01 - 0.02 g/cm/s, the energy absorption is possible for particles of sufficiently large diameters, when the magnetization reversal occurs due to the rotation of nanoparticles as a whole in the viscous mode.

On the other hand, as Fig. 6b shows, at the field amplitude $H_0$ = 300 Oe the dependences of SAR on the particle diameter for dilute assemblies of nanoparticles distributed in liquid and in solid matrix differ significantly. This is the consequence of the fact that at large ac field amplitudes, $H_0 \sim H_k$, the magnetic oscillation mode, which is characterized by an almost rectangular hysteresis loop, is realized in a liquid. According to Fig. 6b, with an increase in particle diameters the SAR of an assembly in liquid reaches very high values, of the order of 1900 W/g, which is comparable to the maximum possible value 2520 W/g for a perfectly rectangular hysteresis loop. Since the particle directors oscillate in the magnetic mode with small amplitude, it is not surprising that the SAR of an assembly in a liquid in magnetic mode is practically independent of viscosity. Figure 6b also shows that the SAR of randomly oriented assembly of nanoparticles in a solid matrix is approximately two times less than that in a liquid, due to the fact that a significant fraction of nanoparticles of the assembly in the solid matrix is not optimally oriented with respect to the ac magnetic field direction.

## 3. Discussion and conclusions

Based on the calculations performed we come to an interesting conclusion that the effect of viscosity on SAR of a dilute assembly of nanoparticles distributed in a liquid is rather limited. Namely, the SAR of the assembly changes inversely with the liquid viscosity only in the developed viscous mode, at small and moderate amplitudes of ac magnetic field, and for nanoparticles of sufficiently large diameters. At large field amplitudes, comparable to the particle anisotropy field, when the magnetic mode is realized, the SAR of assembly in a liquid significantly exceeds the SAR of the same randomly oriented assembly in a solid matrix. But this is a consequence of the orientation of the particle directors in liquid along the ac magnetic field direction. The viscosity of the liquid affects only the time necessary for establishment of the stationary magnetic mode. The latter can be as long as several tens or even hundreds of ac magnetic field periods for particles of large diameters.

Obviously, the SAR of an assembly in a liquid should increase as a result of partial orientation of the particle directors due to the application of external static magnetic field $H_{dc}$ parallel to the ac field direction. Experimental confirmation of this fact was obtained in [46,47]. It is also known that the SAR of an assembly increases significantly upon the formation of one-dimensional dipole-bound chains of nanoparticles in a viscous liquid [48-51]. In this regard, it should be noted that the spontaneous orientation of the particle directors in a viscous fluid, predicted in [31] and confirmed experimentally [38,39], occurs only under the action of ac magnetic field of sufficiently large amplitude, in the absence of external dc magnetic field, and in a sufficiently dilute assembly, neglecting the influence of magnetic dipole interactions between nanoparticles.

It is also worth noting that the calculated SAR values shown in Figs. 1, 4-6 were obtained in this work by numerically integrating low frequency hysteresis loops of assemblies of superparamagnetic nanoparticles according to the well-known [52,53,40] thermodynamic formula, $SAR = M_s f \oint \vec{\alpha} d\vec{H} / \rho$, without invoking any additional hypotheses about the dynamic magnetic susceptibility of the assembly [27], or the relaxation times of nanoparticles [30,54]. Low-frequency hysteresis loops, examples of which are shown in Fig. 2d, were calculated according to the approach described in the methods section. As shown earlier [45], this calculation procedure reproduces quite accurately the known analytical approximations for the relaxation times $\tau_N$ of nanoparticles in a solid matrix. Moreover, it can be shown that it also accurately reproduces the Debye time $\tau_B$ of orientational relaxation of an assembly of nonmagnetic nanoparticles in a viscous liquid. It is worth noting that the low frequency hysteresis loops of the assembly allow direct experimental measurement [3,55,56]. All these facts constitute a solid basis for the approach outlined in this paper. At the same time, the recent attempts [30,54] to introduce the relaxation times, which depend on the ac magnetic field amplitude and frequency seem to us physically unreasonable at sufficiently large amplitudes and frequencies, which are essential for magnetic hyperthermia. Indeed, it is impossible to indicate a direct physical experiment in which such relaxation times could be directly measured.

At the same time, the effective Schliomis relaxation time (2) has a clear physical meaning, since this quantity, at least in principle, allows direct physical measurement. The Shliomis's hypothesis is not trivial. First, it refers only to the process of relaxation of the average magnetic moment of the assembly, and says nothing about the simultaneous orientational relaxation of the nanoparticles of the assembly. In this work it is shown that orientational relaxation in a liquid of finite viscosity occurs with the Debye relaxation time $\tau_B$. Besides, it was shown [57] that in a system with very low viscosity, close to zero, the strong interaction of the rotational and magnetic degrees of freedom of a nanoparticle can lead to new physical effects.

## 4. Methods: Stochastic Landau- Lifshitz and director's equations

In Ref. 31 a system of stochastic equations was proposed to study the coupled dynamics of the unit magnetization vector of the particle $\vec{\alpha}$ and the particle director $\vec{n}$ in a viscous liquid at a finite temperature. The dynamics of the unit magnetization vector is described by the stochastic Landau-Lifshitz equation [44,58-60]

$$\frac{\partial \vec{\alpha}}{\partial t} = -\gamma_1 [\vec{\alpha}, \vec{H}_{ef} + \vec{H}_{th}] - \kappa \gamma_1 [\vec{\alpha}, [\vec{\alpha}, \vec{H}_{ef} + \vec{H}_{th}]]. \tag{7}$$

Here the effective magnetic field is calculated as the derivative of the total particle energy

$$\vec{H}_{ef} = -\frac{\partial W}{V M_s \partial \vec{\alpha}} = \vec{H} + H_k (\vec{\alpha}\vec{n})\vec{n}, \tag{8}$$

where $\vec{H}$ is the vector of a static or variable external magnetic field, $\vec{H}(t) = \vec{H}_0 \cos(\omega t)$. In turn, the dynamics of the nanoparticle director is determined by the stochastic equation

$$\frac{\partial \vec{n}}{\partial t} = G(\vec{\alpha}\vec{n})(\vec{\alpha} - (\vec{\alpha}\vec{n})\vec{n}) - \frac{1}{\xi}[\vec{n}, \vec{N}_{th}], \tag{9}$$

where $\xi = 6\eta V$ is the coefficient of friction of a spherical nanoparticle in a viscous liquid, obtained in the Stokes approximation [61], and the coefficient $G = 2K_1 V/\xi = K_1/3\eta$. Eqs (7), (9) also include the thermal magnetic field $\vec{H}_{th}$ and the fluctuating mechanical torque $\vec{N}_{th}$. In accordance with the fluctuation-dissipation theorem [44, 60], the components of these vectors satisfy the statistical relations

$$\langle H_{th,i}(t)\rangle = 0; \qquad \langle H_{th,i}(t)H_{th,j}(t_1)\rangle = \frac{2k_B T \kappa}{|\gamma_0| M_s V} \delta_{ij} \delta(t - t_1); \tag{10}$$

$$\langle N_{th,i}(t)\rangle = 0; \qquad \langle N_{th,i}(t)N_{th,j}(t_1)\rangle = 2k_B T \xi \delta_{ij} \delta(t - t_1). \tag{11}$$

The method for solving the system of stochastic equations (7), (9) is described in detail in Ref. 31.

**References**
1. Pankhurst, Q. A., Thanh, N. T. K., Jones, S. K. & Dobson, J. Progress in applications of magnetic nanoparticles in biomedicine. *J. Phys. D: Appl. Phys.* **42**, 224001 (2009).
2. Dutz, S. & Hergt, R. Magnetic nanoparticle heating and heat transfer on a microscale: Basic principles, realities and physical limitations of hyperthermia for tumour therapy. *International Journal of Hyperthermia* **29**, 790–800 (2013).
3. Périgo, E. A. *et al.* Fundamentals and advances in magnetic hyperthermia. *Applied Physics Reviews* **2**, 041302 (2015).
4. Silva, A. *et al.* Medical applications of iron oxide nanoparticles. In *Iron Oxides: From Nature to Applications* (ed. Faivre, D.) 425–471 (Wiley, New York, 2016).
5. Blanco-Andujar, C., Teran, F. J. & Ortega, D. Current Outlook and Perspectives on Nanoparticle-Mediated Magnetic Hyperthermia. In *Iron Oxide Nanoparticles for Biomedical Applications* 197–245 (Elsevier, 2018). doi:10.1016/b978-0-08-101925-2.00007-3.
6. Di Corato, R. *et al.* Magnetic hyperthermia efficiency in the cellular environment for different nanoparticle designs. *Biomaterials* **35**, 6400–6411 (2014).
7. Sanz, B. *et al.* In Silico before In Vivo: how to Predict the Heating Efficiency of Magnetic Nanoparticles within the Intracellular Space. *Sci Rep* **6**, (2016).
8. Etheridge, M. L. *et al.* Accounting for biological aggregation in heating and imaging of magnetic nanoparticles. *Technology* **02**, 214–228 (2014).
9. Jeon, S. *et al.* Quantifying intra- and extracellular aggregation of iron oxide nanoparticles and its influence on specific absorption rate. *Nanoscale* **8**, 16053–16064 (2016).
10. Branquinho, L. C. *et al.* Effect of magnetic dipolar interactions on nanoparticle heating efficiency: Implications for cancer hyperthermia. *Sci Rep* **3**, 2887 (2013).
11. Usov, N. A., Serebryakova, O. N. & Tarasov, V. P. Interaction Effects in Assembly of Magnetic Nanoparticles. *Nanoscale Res Lett* **12**, 489 (2017).
12. Soukup, D. *et al*. In Situ Measurement of Magnetization Relaxation of Internalized Nanoparticles in Live Cells. *ACS Nano* **9**, 231–240 (2015).
13. Fortin, J.-P. *et al*. Size-Sorted Anionic Iron Oxide Nanomagnets as Colloidal Mediators for Magnetic Hyperthermia. *J. Am. Chem. Soc.* **129**, 2628–2635 (2007).
14. Mehdaoui, B. *et al.* Large specific absorption rates in the magnetic hyperthermia properties of metallic iron nanocubes. *Journal of Magnetism and Magnetic Materials* **322**, L49–L52 (2010).
15. Guardia, P. *et al.* Water-Soluble Iron Oxide Nanocubes with High Values of Specific Absorption Rate for Cancer Cell Hyperthermia Treatment. *ACS Nano* **6**, 3080–3091 (2012).
16. Boskovic, M. *et al.* Influence of size distribution and field amplitude on specific loss power. *Journal of Applied Physics* **117**, 103903 (2015).
17. Soetaert, F., Kandala, S. K., Bakuzis, A. & Ivkov, R. Experimental estimation and analysis of variance of the measured loss power of magnetic nanoparticles. *Sci Rep* **7**, 6661 (2017).
18. Cabrera, D. *et al.* Unraveling viscosity effects on the hysteresis losses of magnetic nanocubes. *Nanoscale* **9**, 5094–5101 (2017).
19. Nemati, Z. *et al.* Improving the Heating Efficiency of Iron Oxide Nanoparticles by Tuning Their Shape and Size. *J. Phys. Chem. C* **122**, 2367–2381 (2018).
20. Avolio, M. *et al.* In-gel study of the effect of magnetic nanoparticles immobilization on their heating efficiency for application in Magnetic Fluid Hyperthermia. *Journal of Magnetism and Magnetic Materials* **471**, 504–512 (2019).
21. Torres, T. E. *et al.* The relevance of Brownian relaxation as power absorption mechanism in Magnetic Hyperthermia. *Sci Rep* **9**, 3992 (2019).
22. Brezovich, I.A., Low frequency hyperthermia: capacitive and ferromagnetic thermoseed methods. *Med. Phys. Monogr. No 16 Biol. Phys. Clin. Asp. Hyperth.*, 82–110 (1988).
23. Hergt, R. & Dutz, S. Magnetic particle hyperthermia—biophysical limitations of a visionary tumour therapy. *Journal of Magnetism and Magnetic Materials* **311**, 187–192 (2007).
24. Rosensweig, R. E. Heating magnetic fluid with alternating magnetic field. *Journal of Magnetism and Magnetic Materials* **252**, 370–374 (2002).

25. Mamiya, H. & Jeyadevan, B. Hyperthermic effects of dissipative structures of magnetic nanoparticles in large alternating magnetic fields. *Sci Rep* **1**, 157 (2011).
26. Taukulis, R. & Cēbers, A. Coupled stochastic dynamics of magnetic moment and anisotropy axis of a magnetic nanoparticle. *Phys. Rev. E* **86**, 061405 (2012).
27. L. Raikher, Yu. & Stepanov, V. I. Physical aspects of magnetic hyperthermia: Low-frequency ac field absorption in a magnetic colloid. *Journal of Magnetism and Magnetic Materials* **368**, 421–427 (2014).
28. Lima, E., Jr. *et al.* Relaxation time diagram for identifying heat generation mechanisms in magnetic fluid hyperthermia. *J Nanopart Res* **16**, 2791 (2014).
29. Jonasson, C. *et al.* Modelling the effect of different core sizes and magnetic interactions inside magnetic nanoparticles on hyperthermia performance. *Journal of Magnetism and Magnetic Materials* **477**, 198–202 (2019).
30. Yoshida, T., Nakamura, T., Higashi, O. & Enpuku, K. Effect of viscosity on the AC magnetization of magnetic nanoparticles under different AC excitation fields. *Journal of Magnetism and Magnetic Materials* **471**, 334–339 (2019).
31. Usov, N. A. & Liubimov, B. Ya. Dynamics of magnetic nanoparticle in a viscous liquid: Application to magnetic nanoparticle hyperthermia. *Journal of Applied Physics* **112**, 023901 (2012).
32. Usadel, K. D. & Usadel, C. Dynamics of magnetic single domain particles embedded in a viscous liquid. *Journal of Applied Physics* **118**, 234303 (2015).
33. Reeves, D. B. & Weaver, J. B. Combined Néel and Brown rotational Langevin dynamics in magnetic particle imaging, sensing, and therapy. *Appl. Phys. Lett.* **107**, 223106 (2015).
34. Weizenecker, J. The Fokker–Planck equation for coupled Brown–Néel-rotation. *Phys. Med. Biol.* **63**, 035004 (2018).
35. Lyutyy, T. V., Hryshko, O. M. & Kovner, A. A. Power loss for a periodically driven ferromagnetic nanoparticle in a viscous fluid: The finite anisotropy aspects. *Journal of Magnetism and Magnetic Materials* **446**, 87–94 (2018).
36. Engelmann, U. M. *et al.* Predicting size-dependent heating efficiency of magnetic nanoparticles from experiment and stochastic Néel-Brown Langevin simulation. *Journal of Magnetism and Magnetic Materials* **471**, 450–456 (2019).
37. Shasha, C. & Krishnan, K. M. Nonequilibrium Dynamics of Magnetic Nanoparticles with Applications in Biomedicine. *Adv. Mater.* 1904131 (2020) doi:10.1002/adma.201904131.
38. Suwa, M., Uotani, A. & Tsukahara, S. Alignment and small oscillation of superparamagnetic iron oxide nanoparticle in liquid under alternating magnetic field. *Journal of Applied Physics* **125**, 123901 (2019).
39. Suwa, M., Uotani, A. & Tsukahara, S. Magnetic and viscous modes for physical rotation of magnetic nanoparticles in liquid under oscillating magnetic field. *Appl. Phys. Lett.* **116**, 262403 (2020).
40. Usov, N. A. Low frequency hysteresis loops of superparamagnetic nanoparticles with uniaxial anisotropy. *Journal of Applied Physics* **107**, 123909 (2010).
41. Respaud, M. Magnetization process of noninteracting ferromagnetic cobalt nanoparticles in the superparamagnetic regime: Deviation from Langevin law. *Journal of Applied Physics* **86**, 556–561 (1999).
42. Shliomis, M. I. Magnetic Fluids. *Phys.-Usp* **17**, 153 (1974).
43. Debye, P. *Polar Molecules*. (Chemical Catalog, New York, 1929).
44. Brown Jr. W.F. Thermal fluctuations of a single-domain particle. *Phys. Rev.* **130**, 1677-1686 (1963).
45. Usov, N.A. & Grebenshchikov, Yu.B. *Micromagnetics of Small Ferromagnetic Particles*. In *Magnetic nanoparticles*, Ed. Prof. S. P. Gubin (Wiley-VCH, 2009, Chap. 8).
46. Shi, G. *et al.* Enhanced specific loss power from Resovist® achieved by aligning magnetic easy axes of nanoparticles for hyperthermia. *Journal of Magnetism and Magnetic Materials* **473**, 148–154 (2019).
47. Ranoo, S., Lahiri, B. B., Muthukumaran, T. & Philip, J. Enhancement in hyperthermia efficiency under in situ orientation of superparamagnetic iron oxide nanoparticles in dispersions. *Appl. Phys. Lett.* **115**, 043102 (2019).
48. Martinez-Boubeta, C. *et al.* Learning from Nature to Improve the Heat Generation of Iron-Oxide Nanoparticles for Magnetic Hyperthermia Applications. *Sci Rep* **3**, 1652 (2013).
49. Serantes, D. *et al.* Multiplying Magnetic Hyperthermia Response by Nanoparticle Assembling. *J. Phys. Chem. C* **118**, 5927–5934 (2014).
50. Simeonidis, K. *et al.* In-situ particles reorientation during magnetic hyperthermia application: Shape matters twice. *Sci Rep* **6**, 38382 (2016).
51. Asensio, J. M. *et al.* To heat or not to heat: a study of the performances of iron carbide nanoparticles in magnetic heating. *Nanoscale* **11**, 5402–5411 (2019).

52. Landau L.D., Lifshitz E.M., *Electrodynamics of Continuous Media* (Pergamon, New York, 1984).
53. Carrey, J., Mehdaoui, B. & Respaud, M. Simple models for dynamic hysteresis loop calculations of magnetic single-domain nanoparticles: Application to magnetic hyperthermia optimization. *Journal of Applied Physics* **109**, 083921 (2011).
54. Engelmann, U. M. *et al.* Heating efficiency of magnetic nanoparticles decreases with gradual immobilization in hydrogels. *Journal of Magnetism and Magnetic Materials* **471**, 486–494 (2019)
55. Gudoshnikov, S. A. *et al.* AC Magnetic technique to measure specific absorption rate of magnetic nanoparticles. *J. Supercond. Nov. Magn.* **26**, 857–860 (2012).
56. Garaio, E. *et al.* A wide-frequency range AC magnetometer to measure the specific absorption rate in nanoparticles for magnetic hyperthermia. *Journal of Magnetism and Magnetic Materials* **368**, 432–437 (2014)
57. Usov, N. A. & Liubimov, B. Y. Magnetic nanoparticle motion in external magnetic field. *Journal of Magnetism and Magnetic Materials* **385**, 339–346 (2015).
58. García-Palacios, J. L. & Lázaro, F. J. Langevin-dynamics study of the dynamical properties of small magnetic particles. *Phys. Rev. B* **58**, 14937–14958 (1998).
59. Scholz, W., Schrefl, T. & Fidler, J. Micromagnetic simulation of thermally activated switching in fine particles. *J. Magn. Magn. Mater*. **233**, 296-304 (2001).
60. Coffey, W.T., Kalmykov, Yu.P. & Waldron, J.T. *The Langevin Equation*, 2$^{nd}$ ed. (World Scientific, Singapore, 2004).
61. Landau L. D. and Lifshitz E. M., *Fluid Mechanics* (2nd edition, Pergamon Press, 1987).